\documentclass[1p]{elsarticle}
\usepackage[centertags]{amsmath}
\usepackage{amsfonts}
\usepackage{amssymb}
\usepackage{amsthm}

\newcommand{\R}{{\mathbb R}}
\newcommand{\C}{{\mathbb C}}
\newcommand{\be}{\begin{equation}}
\newcommand{\eeq}{\end{equation}}
\newcommand{\bea}{\begin{eqnarray}}
\newcommand{\eea}{\end{eqnarray}}
\newcommand{\ba}{\begin{array}}
\newcommand{\ea}{\end{array}}
\def\nn{\nonumber}
\newcommand{\ft}[2]{{\textstyle\frac{#1}{#2}}}

\newcommand{\ii}{\mathrm{i}}
\newcommand{\ee}{\end{equation} }

\newcommand{\one}{{\rm 1\kern -.9mm l}}

\begin{document}

\title{ Wilson Loops and Chiral Correlators on  Squashed Spheres\tnoteref{t1}}
\tnotetext[t1]{Talk presented by F.Fucito at the conference ``Interactions between Geometry and Physics''\\
17-22 August 2015, Guaruj\'a, Sa\~{o} Paulo, Brasil}

\author[ff]{F. Fucito}
\ead{fucito@roma2.infn.it}
\author[jfm]{J.F. Morales}
\ead{morales@roma2.infn.it}
\author[rp]{and R.Poghossian}
\ead{poghos@yerphi.am}
\address[ff]{Albert Einstein Center for Fundamental Physics (AEC), University of Bern, Sidlerstrasse 5, 3012 Bern, Switzerland\\
and I.N.F.N - sezione di Roma 2, Universit\`a di Roma Tor Vergata, Dipartimento di Fisica, Via della Ricerca Scientifica, I-00133 Roma, Italy}
\address[jfm]{I.N.F.N - sezione di Roma 2 and Universit\`a di Roma Tor Vergata, Dipartimento di Fisica, Via della Ricerca Scientifica, I-00133 Roma, Italy}
\address[rp]{Yerevan Physics Institute, Alikhanian Br. 2, AM-0036 Yerevan, Armenia}

\begin{abstract}
After a very brief recollection of how my scientific collaboration with Ugo started, in this talk I will present some recent results obtained with localization: the deformed gauge theory
partition function $Z(\vec\tau|q)$ and the expectation value of circular Wilson loops $W$ on a
squashed four-sphere will be computed. The partition function is deformed by turning on $\tau_J \,{\rm tr} \, \Phi^J$ interactions  with $\Phi$ the ${\cal N}=2$ superfield.
 For the ${\cal N}=4$ theory SUSY gauge theory exact formulae for $Z$ and $W$ in terms of an underlying $U(N)$  interacting matrix model can be derived thus replacing the free Gaussian model describing the undeformed
${\cal N}=4$ theory.
These results will be then compared with those obtained with the dual CFT according to the AGT correspondence. The interactions introduced previously are in fact related to the insertions of commuting
integrals of motion in the four-point CFT correlator and the chiral correlators are expressed as $\tau$-derivatives of the gauge theory partition function on a finite $\Omega$-background.
\end{abstract}

%
%

%
%

%

\maketitle
\flushbottom
\section{Introduction and summary}

My collaboration with Ugo started around the end of the past century and the beginning of the new one. Before that I had authored few papers on non perturbative results in supersymmetric gauge theories
but I did not know how to treat  the singularities of the moduli space of the gauge connections and I was looking for someone who could help me sort out what I needed out of a vast mathematical literature
in which I did not feel at ease. Pietro Fr\`e, who then was Professor at SISSA adviced me to get in touch with Ugo who, in his words, knew ``everything about moduli spaces of gauge connections''.
So I got in touch with him and we wrote a first paper together \cite{Bruzzo:2000di} in collaboration with A.Tanzini, who later moved to SISSA, and G.Travaglini who were both doing their PhD under
my supervision at that time. In this first work together we showed that the measure one uses to integrate over the moduli space of gauge connections is hyperk\"ahler and that, as a consequence,
the partition function of the ${\cal N}=2$ supersymmetric Yang-Mills gauge theory (SUSY YM from now on) can be written as a form to be evaluated at the boundary of the moduli space. And that was exactly
our problem: the moduli space of gauge connections is singular on the boundary and we did not know how to make sense of that form at such a boundary. The way was pointed out to us first by \cite{Hollowood:2002ds}
and later in \cite{Nekrasov:2002qd}. This latter paper was particularly obscure but, with the help of Ugo, it did not take much to us to rederive the results in our own way and to go beyond \cite{Bruzzo:2002xf}.
In this same period also R.Poghossian, who later became a collaborator of Ugo and mine, came to the same conclusions \cite{Flume:2002az}.

In the years, this line of research has proved to be very fruitful. It would be pointless to give now a detailed account of all of the results in this field but the capacity of these ideas to permeate also other
research fields is extraordinary: string theory, integrable systems, Wilson loops, topological string and supergravity theories, AGT correspondence are all research fields which have benefitted from these ideas.
This only on the physics side and the mathematics counterpart is equally rich! The results I am presenting here stem from this old research and from one of its natural outcome. Besides the partition function, the other
natural observable to compute in an ${\cal N}=2$ SUSY YM theory are correlators of arbitrary powers of scalar fields which are well known to form the so called chiral ring. This is what has been done in \cite{Flume:2004rp}
in which we showed how these variables should be treated in order to compute correlators with localization. It now turns out that this very treatment applies also in the case of the circular Wilson loops
which are some of  the most interesting observables to investigate in  a gauge theory.

In theories with extended supersymmetry, the natural loops to consider are those preserving a fraction of the supersymmetry. The supersymmetric Wilson loop
 typically  involves both the gauge field $A_m$ and its scalar superpartners  $\varphi_i$
\be
W=\langle {\rm tr} \,e^{\cal C} \rangle  \qquad {\rm with} \qquad {\cal C}=\ii \int (A_m \dot x^m+\varphi_i \, \dot y^i) ds
\ee
 Supersymmetry requires $|\dot x|=|\dot y|$. Many such solutions have been found preserving different fractions of the original  supersymmetry \cite{Drukker:1999zq,Zarembo:2002an,Drukker:2007qr}.

As previously mentioned, in this talk, which is largely drawn from \cite{Fucito:2015ofa}, I will focus on circular Wilson loops for theories with ${\cal N}=2,4$ supersymmetry on a squashed $S^4$.
The computations I previously described were carried out on flat space and the partition function or the correlators were dominated by the contribution of instantons or anti-instantons. The Wilson loop is
a real variable and it must keep into account contributions from both instanton and anti-instantons at the same time. This is done \cite{Pestun:2007rz} by multipliying the contribution of the two coordinate patches in which
a sphere is divided along the lines of similar computations on complex manifolds \cite{Nakajima:2003uh,Bruzzo:2009uc}.

Before being treated with localization, circular Wilson loops in ${\cal N}=4$ theory  were first considered in \cite{Erickson:2000af} where it  was conjectured that the perturbative series for $W$ at large N
reduces to a simple counting of ladder diagrams.  The counting of diagrams was extracted from a Gaussian matrix model \cite{Brezin:1977sv}. An exact  formula for $W$ was then proposed in \cite{Drukker:2000rr}
by an explicit evaluation of the Gaussian matrix model partition function to all orders in $1/N$.

As previously mentioned, the conjectured formula for the  circular Wilson loop in the ${\cal N}=4$ theory was later proved in \cite{Pestun:2007rz}  using  localization.
 In that reference, the partition function of a general ${\cal N}=2$  gauge theory  on a round sphere $S^4$ was shown to be given by the integral $\int da |Z(a,\tau)|^2$ with $Z(a,\tau)$ the gauge partition
function on $\R^4$. Here $Z$ is evaluated  on an $\Omega$-background such that $\epsilon_1=\epsilon_2=\epsilon$ are finite, so the gauge theory lives effectively on a non-trivial
gravitational background \cite{Nekrasov:2003rj,Billo:2006jm}. On the other hand the expectation value of a circular Wilson loop on $S^4$ was shown to be given by the same integral with the insertion of
  the phase ${\rm tr} \,e^{\frac{2\pi i a}{\epsilon}}$ inside the integral.  The ${\cal N}=4$ Gaussian matrix model was then recovered from  the case of ${\cal N}=2^*$ theory
in the limit where the mass of the adjoint hypermultiplet is sent to 0.

  It is natural to ask how these results are modified when the YM action is deformed or the spacetime is changed. An interesting spacetime to consider is that
  of a squashed sphere \cite{Hama:2012bg}. A motivation for considering four-dimensional gauge theories on squashed spheres comes from the AGT correspondence that relates the partition function
of ${\cal N}=2$ supersymmetric theories to correlators in a dual two-dimensional CFT with the squeezing
parameter  $\epsilon_1/\epsilon_2$ parametrizing the central charge of the CFT  \cite{Gaiotto:2009we,Alday:2009aq,Drukker:2009id}.
  In \cite{Alday:2009fs}, it was proposed that the expectation value of a circular Wilson loop oriented along, let us say the first plane, on the squashed sphere is given by the
  insertion of  the phase ${\rm tr} \,e^{2\pi i a\over \epsilon_1}$ in the partition function integral.
  Here I will prove this formula using localization and extend it to the case of a general ${\cal N}=2$ theory with prepotential
  ${\cal F}_{\rm class}=\tau_J\, {\rm tr}\,\Phi^J$.
  This class of gauge theories were first introduced in \cite{Losev:2003py}  and the deformed partition function  $Z(a,\vec\tau)$  was computed  in \cite{Nakajima:2003uh} (see also \cite{Nekrasov:2009rc})
and  related to the generating function of Gromov-Witten invariants for certain complex manifolds.

   In this talk I will present a "physical derivation" of the deformed partition function  based on localization and the results of \cite{Flume:2004rp}and apply it to the study of circular Wilson loops on squashed spheres.
The crucial observation is that the circular Wilson loop ${\cal C}$  is nothing but the lowest component of the  equivariant superfield
   ${\cal F}$ defining the field strength of a connection on  the so called universal moduli space, the moduli space of instantons times the space time. Indeed in presence of an $\Omega$ background the
zero mode solutions to the equations of motion get deformed and the lowest component of ${\cal F}$ reduces to
 $\tilde\varphi =\varphi+\delta_\xi x^m\, A_m$ with $\delta_\xi x^m$ a rotation of the spacetime coordinates along the Cartan of the Lorentz group.  The same combination appears on
the circular Wilson loop  ${\cal C}={2\pi \ii n_1 \over \epsilon_1} \tilde\varphi $, so the expectation value of the Wilson loop
 can be related to the  $\Omega$-deformed version of the corresponding chiral correlator. Moreover using the fact that Wilson loops are allowed only on spheres with  rational squeezing parameter $\epsilon_1/\epsilon_2$,
an explicit evaluation of  $\tilde\varphi$ shows that
 $\left\langle  e^{\cal C} \right\rangle =\left\langle e^{ {2\pi \ii n_1 \over \epsilon_1} a} \right\rangle $, so all instanton corrections to  ${\cal C}$ (but not to the correlator) cancel after exponentiation !  The
same cancellation was observed in \cite{Pestun:2007rz} for the case of the round sphere.
   Similarly ${\rm tr}\,\Phi^J$ interactions can be related to the higher component of ${\rm tr}\, {\cal F}^J$,  and their expectation values are again given by the
   deformed chiral correlators up to some super-volume factors.

    The case of the deformed ${\cal N}=4$ theory is particularly interesting. $\tau_J$-interactions break supersymmetry down
    to ${\cal N}=2$  but surprisingly the resulting theory is far more simpler than any other ${\cal N}=2$ theory one can think of.
       For instance instanton contributions to the partition function and circular Wilson loops can be shown to cancel in
       the deformed theory more in the same way than in the  maximally supersymmetric theory. As a consequence the gauge partition function and the expectation values of circular Wilson loops  are
described by an effective interacting matrix model.
  More precisely, the Wilson loop is given by the integral
    \be
    \left\langle \, {\rm tr} \, e^{ \cal C} \right\rangle ={1\over Z} \int  d^Na\, \Delta(a)  \, {\rm tr} \,e^{ 2\pi a \over \epsilon_1}\, e^{- N\,  V(\vec\tau,a)}
    \ee
with $\Delta(a)$ the Vandermonde $U(N)$ determinant  and $V(\vec\tau,a)$ a potential codifying the $\tau_J$-interactions.
The matrix model integral effectively counts the number of Feynman diagrams with propagators ending on the Wilson loop, and internal
J-point vertices weighted by $\tau_J$.  The case with only quartic interactions will be worked out
in full details.

    Exploiting the AGT correspondence between 4d gauge theories and 2d CFT's,  the chiral correlators $\langle \, {\rm tr} \, \tilde\varphi^J \rangle$ in the gauge theory can be related to insertions of the
integrals of motions $I_J$ into the four-point CFT correlator computing the gauge theory partition function \cite{Nekrasov:2009rc,Bonelli:2014iza}.
The computation of such deformed correlators is feasible and it leads to relations which are the analogue
of the chiral ring relations for ${\cal N}=2$ gauge theories
\cite{Cachazo:2002ry} for a finite $\Omega$ background. The explicit computation will be carried out in the $SU(2)$ theory with four fundamentals
for the simplest chiral correlators and the result will match with the dual  correlators in the Liouville theory.

In the so called Nekrasov-Shatshvili limit in which one of the two deformation parameters of the $\Omega$ background is sent to zero an
$\epsilon$-deformed version of the Seiberg-Witten type curve is available \cite{Poghossian:2010pn,Fucito:2011pn} from which we one can extract the full set of chiral ring relations
and compare against the results presented here \cite{Fucito:2015ofa}.

\section{The gauge partition function on $\R^4$}

The action of a general ${\cal N}=2$  supersymmetric gauge theory on $\R^4$ is specified by a single holomorphic function,  {\it the prepotential},
${\cal F}_{\rm class}(\Phi)$  of the ${\cal N}=2$ vector multiplet superfield $\Phi$. Explicitly
 \be
S_{\rm class}=\left[ {1\over 4 \pi^2} \int d^4x d^4 \theta \, {\cal F}_{\rm class}(\Phi)+{\rm h.c.}\right]+\ldots
\ee
 with $(x_{m},\theta_{\alpha\dot a})$ denoting the super-space coordinates and the lower dots denoting couplings to hypermultiplets. I am interested in the Coulomb branch of the gauge theory where
the scalar $\varphi$ in the vector multiplet $\Phi$ has a non-trivial vacuum expectation value.
In this branch, only the vector multiplet prepotential is corrected by quantum effects.
 Taking the classical prepotential of the general form
\be
{\cal F}_{\rm class}(\Phi)= \sum_{J=2}^p  {2\pi  \ii \tau_J\over J!} \, {\rm tr}\,   \Phi^J   \label{fphi}
\ee
for some integer $p$ and
\bea
\Phi &=& \varphi+ \lambda_m \theta^m+ \ft12 \,F_{mn} \theta^m \theta^n +\ldots    \label{phisuper}
\eea
  Here and in the rest of the paper I use the "twisted" fermionic variables $\theta^m=\ft12 \sigma^{m\alpha\dot a} \theta_{\alpha\dot a}$ obtained by identifying
 internal  $\dot a$ and Lorenz  $\dot \alpha$ spinors indices.
 The gauge theory following from (\ref{fphi}) can be seen as a deformation of the standard renormalizable gauge theory
where the gauge coupling is replaced by a function of the scalar field  $\varphi$ and the super symmetrically related  couplings $F\lambda^2$ and $\lambda^4$ are included.
The standard theory with prepotential  ${\cal F}_{\rm class}(\Phi)= \pi \ii \tau  {\rm tr}\,   \Phi^2$ is recovered after setting  $\tau_{J>2} $ to zero.

I will refer to the partition function $Z(\vec \tau ,q)$ as the ``deformed partition function".
 There are two main contributions to the partition function $Z_{\rm one-loop}$ and $Z_{\rm tree+inst}(\vec \tau,q)$ coming from the fluctuations
 of the gauge theory fields around the trivial and the instanton vacua.  I notice that only the  latter one depends
 on the couplings $\tau_J$ since  $Z_{\rm one-loop}$ is given by a one-loop vacuum amplitude.  On the other hand $Z_{\rm tree+inst}(\vec \tau,q)$
 can be written in terms of an integral over the instanton moduli space that can be computed with the help of localization.
To do so one needs to introduce an equivariant supercharge $Q_\xi$ and an equivariant vector superfield ${\cal F}$.
Due to a limited number of pages I will not show this treatment in detail and refer the reader to \cite{Fucito:2015ofa}.

Finally the field   ${\cal F}$ can then be viewed as the equivariant version of the ${\cal N}=2$ superfield $\Phi$.
Moreover it is $Q_\xi$-invariant up to a symmetry rotation. This implies that any invariant function made out of ${\cal F}$ is $Q_\xi$-closed.
  In particular the Yang-Mills action (\ref{fphi}) in the instanton background  can be written in the explicit  $Q_\xi$-invariant form
  \be
   S_{\rm inst}(\vec \tau) = \sum_{J=2}^p {\ii \tau_J\over 2\pi J!} \,  \int_{\C^2} {\rm tr}\, {\cal F}^J     \label{fj}
   \ee
For concreteness, from now on, I will focus on conformal gauge theories with gauge group $U(N)$  and fundamental or adjoint matter.

\subsection{The deformed gauge partition function}

The gauge partition function on $\R^4$ can be written  as the product of the one-loop, tree level and instanton contributions
    \be
     Z(\vec \tau)=   Z_{\rm one-loop}  \,  Z_{\rm inst+tree}(\vec\tau)  \label{ztotal0}
     \ee
      The one-loop partition function $Z_{\rm one-loop}$   is given by
    \be
Z_{\rm one-loop }=  Z^{\rm gauge}_{\rm one-loop }\,  Z^{\rm matter}_{\rm one-loop }   \label{zoneloop0}
\ee
with \cite{Alday:2009aq,Fucito:2013fba}
\bea
Z^{\rm gauge}_{\rm one-loop } &=&  \prod_{u < v}^N      \Gamma_2(a_{uv} )^{-1}  \Gamma_2(a_{uv}+\epsilon )^{-1}      \nn\\
Z^{\rm matter}_{\rm one-loop } &=&  \left\{
\begin{array}{ll}
 \prod_{u<v}^N   \Gamma_2(a_{uv}-m)  \Gamma_2(a_{uv}+m+\epsilon)       &   {\rm adj.}   \\
  \prod_{u,v=1}^N  \Gamma_2( a_v-\bar{m}_{u}  ) \Gamma_2( a_u-m_{v+N}+\epsilon)   & {\rm fund.}    \\
\end{array}
\right.
   \label{zloop}
 \eea
 Here $\Gamma_2(x)$ is the Barnes  double gamma function\footnote{  The  Barnes  double gamma function can be thought of as a regularization of the infinite product (for $\epsilon_1>0,\epsilon_2>0$)
 \be
\Gamma_2(x) =\prod_{i,j=0}^\infty   \left( {\Lambda \over x+i \epsilon_1+j\epsilon_2}\right)
\ee
} and $\epsilon=\epsilon_1+\epsilon_2$.

 The instanton partition function is defined by the moduli space integral
   \be
 Z_{\rm inst+tree}( \tau) =  \sum_{k=0}^\infty   \, \int d\mathfrak M_{k,N} \,e^{-S_{\rm inst}( \vec\tau) }    \label{zinst}
 \ee
 with the integral running over the instanton moduli space for a given $k$. The action $S_{\rm inst}$ and the integral over the instanton moduli
 spaces can be evaluated with the help of localization leading to
\be
S_{\rm inst}(\vec \tau)=  \sum_{J=2}^p  {\ii \tau_J\over 2 \pi J!}  \, \int_{\C^2} {\rm tr}\, {\cal F}^J     =
{2\pi  \ii \over  \epsilon_1 \epsilon_2}  \, \sum_{J=2}^p  {\tau_J\over J!}  \,  {\rm tr}\, {\tilde \varphi}(0)^J    \label{sphitilde}
   \ee
    The symmetry group of the integral is $ U(N)\times U(N_f)\times U(k)\times SO(4)$. Let me parametrize
     the Cartan element by $a_u, m_i, \chi_I, \epsilon_{1,2}$. More precisely, $a_u$ parametrizes the eigenvalues of the vacuum expectation value
     matrix $\langle \varphi \rangle$, $m_i$ the masses of the fundamental or adjoint fields, $\chi_I$ the
     Cartan of U(k) and  $\epsilon_{1,2}$ are Lorentz breaking parameters that deform the
    $\R^4$ spacetime geometry. For a gauge theory on flat space $\epsilon_{1,2}$ should be sent to zero at the end of the computation. For finite $\epsilon$'s the integral describes the partition function
     on a non-trivial gravitational  background, the so called $\Omega$-bakcground. The fixed points of of the symmetries of the theory are isolated
   and in one-to-one correspondence with N-tuples of Young tableaux $Y=(Y_1,\ldots Y_N)$ with a total number of $k$ boxes. The Young tableaux $Y$
   specify the U(k) Cartan elements $\chi_I$. Explicitly
   \be
   \chi_{(i,j)}=a_u+(i-1) \epsilon_1+(j-1) \epsilon_2      \qquad   (i,j)\in Y_u    \label{chiij}
   \ee
   In our conventions $\epsilon_1,\epsilon_2$ are pure real numbers.
  One then finds for the moduli space integral
    \be
    Z_{ \rm inst+tree}( \tau) =\sum_{Y}    Z_Y\, e^{-S_Y(\vec \tau)}   \label{zint}
\ee
with $Z_Y$ the inverse determinant of $\delta_\xi$ and $S_Y(\vec \tau)$ the 0-form of the instanton action (\ref{sphitilde}) evaluated at the fixed point.
 For $Z_Y$ one finds with \cite{Flume:2002az,Bruzzo:2002xf,Fucito:2013fba} $Z_Y= { 1 \over {\rm det} \, \delta_\xi }\Big|_{Y}=   Z_Y^{\rm gauge} \, Z_Y^{\rm matter}$
and
\bea
Z_Y^{\rm gauge} &=& \prod_{u,v}^N    Z_{Y_u,Y_v}(a_{uv})^{-1}    \nn\\
Z_Y^{\rm matter} &=&   \left\{
\begin{array}{ll}
   \prod_{u,v}^N   Z_{Y_u,Y_v}(a_{uv}+m)     &   {\rm adjoint}    \\
  \prod_{u,v=1}^N    Z_{\emptyset,Y_v}( \bar{m}_{u}-a_v  ) \, Z_{Y_u,\emptyset}(a_u-m_{v+N} ) &  {\rm fund}    \\
\end{array}
\right.
  \label{zinst0}
\eea
and
\bea
  Z_{Y_u,Y_v}(x)  &=& \prod_{ (i,j)\in Y_u}
      (x-\epsilon_1(k_{vj}-i)+\epsilon_2 \, (1+\tilde k_{ui}-j)) \nn\\
      \times && \prod_{ (i,j)\in Y_v} (x+\epsilon_1(1+ k_{uj}-i)-\epsilon_2 \, (\tilde k_{vi}-j))
      \label{zalbe2}
 \eea
 Here $(i,j)$ run over  rows and columns respectively of  the given Young tableaux,
$\{ k_{uj} \} $  and $ \{ \tilde k_{ui} \}$ are positive integers
giving the length of the rows and columns respectively of the tableau $Y_u$.
 I remark that  (\ref{zloop}) and (\ref{zinst0}) are not
symmetric under the exchange  of fundamental $\bar m_i$ and  anti-fundamental masses $\bar m_j$ but a totally symmetric form
under the exchange $\bar m_i \leftrightarrow \bar m_j$ can be obtained by replacing
 \be
m_{u+N} \to  \bar m_{u+N}+\epsilon
 \ee
   Finally, the contribution of the deformed Yang-Mill action  at the fixed point $Y$ reduces to
  \be
S_Y(\vec \tau) ={2\pi \ii \over \epsilon_1 \epsilon_2}  \, \sum_{J=2}^p  {\tau_J\over J!} \,   {\cal O}_{J,Y}
\ee
with \cite{Losev:2003py,Flume:2004rp}
 \be
{\cal O}_{J,Y}=   \sum_u\left( a_u^J
 -   \sum_{(i,j)\in Y_u}    \left[ \chi_{(i,j)}^J+(\chi_{(i,j)}+\epsilon)^J-  (\chi_{(i,j)}+\epsilon_1)^J-(\chi_{(i,j)}+\epsilon_2)^J    \right]  \right)
\label{varphifix}
\ee
 The deformed partition functions can then be written as
    \be
   Z_{\rm inst+tree} (\vec \tau)=  \sum_{Y}    Z_Y(\vec \tau) =  \sum_{Y}    Z_Y\,  {\rm exp}   \left( - {2\pi \ii \over \epsilon_1 \epsilon_2}  \, \sum_{J=2}^p {\tau_J\over J!} \,    {\cal O}_{J,Y}  \right)   \label{zdef}
    \ee
    with   ${\cal O}_{J,Y} $ given by (\ref{varphifix}). The gauge prepotential ${\cal F}_{\rm eff}$
 is then identified with the free energy associated to the deformed instanton partition function
  \be
 {\cal F}_{\rm eff} ( \vec \tau )=-   \epsilon_1 \, \epsilon_2\, \ln Z(\vec \tau)   \label{fdef}
 \ee
 Now
\be
 \langle  {\rm tr}\,  \tilde\varphi^J  \rangle =  {1\over Z_{\rm inst+tree}  } \sum_Y Z_Y(\vec\tau)\,   {\cal O}_{J,Y}
 \ee
  or equivalently from (\ref{zdef})
  \be
 {1\over J!}  \langle  {\rm tr}\,  \tilde\varphi^J \rangle
 ={\ii \epsilon_1 \epsilon_2 \over 2\pi } \, \partial_{\tau_J} \ln Z(\vec{\tau}) \label{phij}
 \ee
The deformed partition function is the generating functional of the general multi-trace chiral correlators
   \be
    \langle  {\rm tr}\,  \tilde\varphi^{J_1} \, {\rm tr}\,  \tilde\varphi^{J_2} \,\ldots  \rangle_{\rm undeformed}
   \ee
   in the undeformed theory.

 \section{The gauge theory on $S^4$ }

  The gauge partition function on $S^4$ is given by the integral  \cite{Pestun:2007rz}
    \be
 Z_{S^4} (\vec \tau)  =c\,   \int_{\gamma}  d^N a \,   |  Z_{\rm one-loop}(  a )\,  Z_{\rm tree+inst} ( a ,\vec \tau) |^2\label{ZS4}
 \ee
 with $d^N a=\prod_u d a_u $ {}\footnote{Notice that  in our conventions the Vandermonde determinant is reabsorbed into the one-loop
 determinant $ Z_{\rm one-loop}$. } and the integral running along the imaginary axis.
The partition functions  $Z_{\rm one-loop}$ and $Z_{\rm inst}$ are given by (\ref{zloop}) and (\ref{zinst0}) with vevs and masses taken
in the domains
\be
a_u =\in \ii \R  \qquad   \qquad   m,\epsilon_\ell \in \R   \qquad   m_u={\epsilon\over 2}+ \ii \R   \qquad    \bar{m}_u=-{\epsilon\over 2} + \ii \R
\ee
These domains are chosen such that complex conjugate of  the one-loop partition function (\ref{zloop}) is given by the same formula with
$\Gamma_2(x)$ replaced by $\Gamma_2(\epsilon-x)$.
Here I use units where $\epsilon_1\epsilon_2=1$ and write $\epsilon_1=\epsilon_2^{-1}=b$. In these units the one-loop partition function  becomes
\bea
|Z^{\rm gauge}_{\rm one-loop }|^2 &=&  \prod_{u < v}^N      \Upsilon( a_{uv} )  \Upsilon(-a_{uv} )      \nn\\
|Z^{\rm matter}_{\rm one-loop }|^2 &=&  \left\{
\begin{array}{ll}
 \prod_{u,v}^N      \Upsilon(  a_{uv}-m)^{-1}     &   {\rm adj.}   \\
  \prod_{u,v=1}^N  \Upsilon( a_v-\bar{m}_{u}  )^{-1} \Upsilon(m_{v+N}- a_u)^{-1}   & {\rm fund.}    \\
\end{array}
\right.
   \label{zloops4}
 \eea
 with
\be
\Upsilon(x)={1\over \Gamma_2\left(x|b,\ft{1}{b}\right)\Gamma_2\left(b+\ft{1}{b} -x|b,\ft{1}{b}\right)  }
\ee
 $\Upsilon$ is an entire function  satisfying $\Upsilon(x)=\Upsilon(\epsilon-x)$. It has an infinite number of single zeros at $x=-m b-n/b$ and $x=(m+1)b+(n+1)b$
 for $m,n\geq 0$ integers.
  Finally the normalization  $c$ has been fixed for later convenience to be\footnote{This normalization is chosen in such a way that the partition function and its AGT dual
  correlator precisely match. }
  \be
  c=q^{ {1\over 2} m_3(m_3+\epsilon)+ {1\over 2} m_4(m_4+\epsilon) }  \label{ccc}
  \ee

  \subsection{ Wilson loops  }

  A supersymmetric Wilson loop  is defined by the line integral
   \be
{\cal C}= \ii \int_0^L (A_m \,\dot{x}^m + |\dot{x}| \, \varphi_1 ) ds  \label{cc}
   \ee
  with  $\varphi_1=\ft12(\varphi-\varphi^\dagger)$  and $A_m$ taken to be anti-hermitian matrices.
  Using complex coordinates $x^m=(z_1,z_2,\bar z_1, \bar z_2)$ and consider a circular Wilson loop
  defined by the path
   \be
  z_\ell (s)= r_\ell \, e^{ \ii \epsilon_\ell s}     \label{solc}
  \ee
  with
   \be
  L= {2\pi  n_1\over \epsilon_1}={2\pi n_2\over \epsilon_2} \label{epsratio}
   \ee
  The condition (\ref{epsratio}) ensures that the path is closed and it can be satisfied if and only if
  the ratio $\epsilon_1/\epsilon_2$ is rational.
  Moreover, taking $r_{1,2}$ satisfying
   \be
   |\dot x|^2= \epsilon_1^2 |r_1|^2+ \epsilon_2^2|r_2|^2=1
  \label{solc1}
  \ee
   one finds
   \be
  \dot x^m =( \ii \epsilon_1 z_1 ,  \ii \epsilon_2 z_2,  -\ii \epsilon_1 \bar z_1 ,  -\ii \epsilon_2 \bar z_2)= \delta_\xi x^m
   \ee
and the Wilson loop can be written in the suggestive form
        \be
    {\cal C}= \ii \int_0^L \left( A_m \, \delta_\xi  x^m +   \varphi_1 \right) ds  =\ft{\ii}{2} \int_0^L \tilde{\varphi}(s) ds+{\rm h.c.}
    \ee
   From this the first $\tau_J$-correction  to the Wilson loop expectation value is computed by the correlator
 \be
  {\partial \over \partial \tau_J} \left \langle {\rm tr}\,   e^{ \cal C }  \right\rangle_{S^4}   = {2 \pi \ii \over \epsilon_1 \epsilon_2 \, J!}
  \left \langle {\rm tr}\,   e^{ \cal C } \,   {\rm tr}\,  \tilde\varphi^{J}  \right\rangle_{S^4,\rm undef.}   \label{deltaw}
 \ee

 \subsection{The  ${\cal N}=4$ deformed theory}

 In this section I consider the effect of turning on $\tau_J$-interactions on the simplest theory at our disposal: the ${\cal N}=4$ theory.  $\tau_J$-deformations break ${\cal N}=4$ supersymmetry
down to ${\cal N}=2$ by including self-interactions for one of
the three chiral multiplets. The resulting theory is surprisingly simple, it exhibits a perfect  cancellation of instanton contributions to the gauge partition  function.   This is in contrast with the  case of ${\cal N}=2^*$ theory where the ${\cal N}=4$ symmetry is broken by giving mass to
the adjoint hypermultiplet spoiling the  balance between the instanton corrections coming from gauge and matter multiplets.

As in the undeformed case, the ${\cal N}=4$ deformed theory will be defined as a
the limit of the ${\cal N}=2^*$ deformed theory where the mass of the
adjoint hypermultiplet is sent to ``zero" \cite{Pestun:2007rz}.  More precisely, the
points where the ${\cal N}=4$ is restored will be identified with the zeros of the instanton partition function in the $m$-plane. It is easy to see that they are located at
 $m=-\epsilon_1$ or $m=-\epsilon_2$. Indeed,  for any choice of $\tau_J$ and
instanton number the instanton partition function (\ref{zinst0}) can be seen to be always proportional to $(m+\epsilon_1)(m+\epsilon_2)$ with the two factors coming from  the
contributions  to $Z_{Y_u Y_u}(m) $ of the  Young tableaux boxes $(i,j)=(k_{uj},\tilde k_{ui} )\in Y_u$.
 Notice that these two points on the $m$-plane coincide in the case of the round sphere where $\epsilon_1=\epsilon_2=\ft{\epsilon}{2}$.
 and the symmetric point $m=-\ft{\epsilon}{2}$ is unique ($m=0$ in the conventions of \cite{Pestun:2007rz}). For concreteness here we take $m=-\epsilon_1$.

The one-loop partition function (\ref{zoneloop0}) also drastically simplifies  at $m=-\epsilon_1$.
Indeed using the double Gamma function identity
\bea
\Gamma_2(x+\epsilon_1) \,\Gamma_2(x+\epsilon_2)=x\, \Gamma_2(x)\, \Gamma_2(x+\epsilon)
\label{id}
\eea
 one finds
  \be
  |Z_{\rm oneloop}|^2 =\Delta(a)=\prod_{u\neq v} a_{uv}
 \ee
 with $\Delta(a)$ the Vandermonde determinant describing the $U(N)$ measure.
 The gauge partition function reduces to the $U(N)$ matrix model integrals
  \be
  Z=  \int d^Na \,\Delta(a) \, e^{- N\,  V(a,\vec\tau)}    \label{zws4}
 \ee
 with the integral over $a_u$ now running along the real line and the potential defined by
 \be
 V(a,\vec\tau)=    {2\pi \ii  \over \epsilon_1 \epsilon_2\, N}  \, \sum_{J=2}^p   {\tau_{J}   \over J!} \,   {\rm tr} (\ii a)^J+{\rm h.c.}
 \label{v1}
 \ee
On the other hand the expectation value of a circular Wilson loop is given by
 \be
 W ={1\over Z} \int d^Na \,\Delta(a) \, {\rm tr} \,e^{2\pi  a \over \epsilon_1}\, e^{- N\,  V(a,\vec\tau)}    \label{ws4}
\ee
 Notice that unlike in the ${\cal N}=4$ theory, in presence of $\tau_J$-interactions the matrix model underlying the theory is no-longer
 Gaussian but interacting. The integrals (\ref{zws4}) and (\ref{ws4}) count now  diagrams involving not only propagators but also $J$-point vertices.
 Luckily enough, the underlying matrix models have been extensively studied in the literature.
    In the large $N$ limit integrals (\ref{zws4}) and (\ref{ws4}) have been explicitly evaluated by saddle point methods \cite{Brezin:1977sv}. One defines the resolvent
  \be
  w(x)={1\over N} \left\langle {\rm tr}\, {1\over x-a} \right\rangle={1\over N} \left\langle \, \sum_{u=1}^N {1\over x-a_u} \right\rangle
  \ee
   A simple algebra shows that $w(x)$ defined like this satisfies a quadratic equation with solution
     \be
     w(x)=\ft12\left(  V'(x)-\sqrt{V'(x)-4 f_{p-2}(x) } \right) \label{wxg}
     \ee
where $f_{p-2}(x)$ a polynomial of order $p-2$ determined by the condition that $w(x)\approx {1\over x}$ at large $|x|$.
 Notice that $w(x)$ has a discontinuity along the cuts defined by the zeroes of the square root, so
 \be
w(x\pm \ii 0) =  \int_{\cal S} {\rho(y) dy\over x-y\pm \ii 0}=\ft12 \, V'(x) \pm \pi \, \ii\, \rho(x)
\label{densita}\ee
with $\cal S$ the union of the cuts and $\rho(x)$ the density
 \be
 \rho(x)={1\over N}\sum_u \delta(x-a_u)
 \ee
  It is often enough to assume the presence of a single cut and look for $w(x)$ in the form
   \be
   w(x)=\ft12 V'(x) -Q_{p-2}(x) \sqrt{ (x-b_1)(x-b_2) }  \label{wq}
   \ee
   with $Q_{p-2}(x)$ a polynomial of order $p-2$.  Indeed the number of unknown variables in $\{ Q_{p-2} (x), b_1, b_2 \}$ is $p+1$, matching
   the number of equations coming from requiring that $w(x)\approx {1\over x}$ for large $|x|$ and therefore $w(x)$ is fully determined.

   As an example, let us consider the quartic potential
   \be
   V(a)={1\over 2\lambda} \, a^2+g_4 \, a^4  \label{v2}
   \ee
with
\be
\lambda={ N\epsilon_1\epsilon_2 \over 4\pi ({\rm Im}\tau_2)} ={g^2_{YM}N\epsilon_1\epsilon_2 \over 16\pi^2}  \qquad ~~~~~~ g_4=-{ 4\pi {\rm Im}(\tau_4)\over 4! N\,\epsilon_1\epsilon_2}   \label{lll}
\ee
Since $V(a)=V(-a)$ one can take $b_2=-b_1=2b$ and $Q(x)=c_0+c_2 x^2$ in (\ref{wq}).
     Requiring $w(x)\approx {1\over x}$ for large $|x|$ one finds \cite{Brezin:1977sv}
   \be
   w(x)={1\over 2\lambda} \left(  x+4\, x^3\, g_4 \, \lambda-(1+4\,\lambda \,g_4  (x^2+2 b^2))\sqrt{x^2-4 b^2}     \right)  \label{wx}
   \ee
   with
   \be
   b^2={ \sqrt{ 1+48 \lambda^2 g_4}-1 \over 24 \lambda g_4}\qquad \Rightarrow\qquad     \lambda = \frac{b^2}{1-12g_4b^4} \label{bsol}
   \ee
  From (\ref{densita}) one can extract the normalized density of eigenvalues $\rho(x)$, which in our case turns out to be
 \be
\rho(x)=\frac{\left[(1+8\lambda g_4 b^2)+4\lambda g_4 x^2\right]\sqrt{x^2-4 b^2}}{2\pi b^2(1+1\lambda g_4b^2)}
\label{densitanorm}
\ee
Now
\bea
<W>&=&\int_{-2b}^{2b}\rho(x)e^\frac{x}{2}=2\left[\frac{I_1(b)}{b}-4g_4b^3I_1(b)+16g_4b^4\frac{\partial^2}{\partial b^2}\left(\frac{I_1(b)}{b}\right)\right]\nonumber\\
&=&{2\, \over\sqrt{ \lambda }}
\left[   I_1 \left(\sqrt{ \lambda } \right)  + 12 \, g_4\, b^4\,  I_3\left(\sqrt{ \lambda } \right)  \right]\eea
with $n_1n_2=1$ and $\epsilon_1\epsilon_2=16\pi^2$.

\section{ AGT duality:  chiral correlators vs integrals of motion  }

\subsection{The CFT side}

The AGT correspondence \cite{Alday:2009aq} relates ${\cal N}=2$ supersymmetric gauge theories in four dimensions to
 two dimensional CFTs.  According to this correspondence the instanton partition function for
 the SU(2) gauge theory with four fundamentals is related to four-point  correlators
 in Liouville theory. In this section the study the CFT side of the duality is put forward and in the next I will study the gauge side of it and show that the chiral correlators $\langle {\rm tr} \tilde \varphi^n \rangle$
in the gauge theory are reproduced  by the same four-point correlators
in Liouville theory with the insertion of  the integrals of motion $I_n$ introduced in \cite{Alba:2010qc}.

 \subsubsection{The conformal field theory}

 Here I follow \cite{Alba:2010qc} for details and further references.
 The symmetry algebra of Liouville theory is the tensor product of a Virasoro and a Heisenberg algebra with commutation relations
 \bea
 \left[ L_m,L_n \right] &=&(m-n)\, L_{m+n}+{c\over 12}(m^3-m)\,\delta_{m+n,0} \nn\\
 \left[ a_m,a_n \right] &=& {m\over 2} \delta_{m+n}  \qquad \left[ L_m ,a_n\right]=0
\eea
The central charge c is parametrized by
\be
c=1+6 \, Q^2    \qquad {\rm where}  \qquad Q=b+{1\over b}
\ee
The primary fields $V_\alpha$ are defined as
\be
V_\alpha(z) = {\cal V}^{\rm vir}_\alpha (z)  \, {\cal V}^{\rm heis}_\alpha (z) \,
\ee
with ${\cal V}^{\rm vir}_\alpha$ a primary field of the Virasoro algebra with dimension
$\Delta(\alpha)=\alpha(Q-\alpha)$ and
\be
 {\cal V}^{\rm heis}_\alpha (z) =e^{2 i (\alpha-Q) \sum_{n<0} {a_n\over n} z^{-n} } \,e^{2 i \alpha \sum_{n>0} {a_n\over n} z^{-n} }
 \ee
 The commutation relations of the field $V_\alpha $ and the generators $L_m$, $a_n$ are
 \bea
  \left[ L_m, V_\alpha (z) \right] &=&
{\cal V}^{\rm heis}_\alpha (z)
\left( z^{m+1}\partial_z+(m+1) \Delta(\alpha)\, z^m\right) \, {\cal V}^{\rm vir}_\alpha (z) \nn\\
   \left[ a_n, V_\alpha(z) \right] &=&
\left\{
\begin{array}{cc}
 {\rm i} \, (Q-\alpha) \, z^n \,   V_\alpha (z)   &   {\rm for} \qquad n>0 \\
  - {\rm i} \, \alpha \, z^n \,   V_\alpha (z)   &   {\rm for} \qquad n<0 \\
\end{array}
\right.
   \label{commrel}
 \eea
  The Fock space is obtained by acting with $L_n,a_n$ with $n<0$ on a vacuum $|0\rangle$ defined by
 \be
 L_m |0\rangle=a_n |0\rangle=0    \qquad {\rm for} \qquad m \geq -1,\quad n>0
 \ee
 Primary states $|\alpha\rangle$ and $\langle \alpha | $ are obtained by acting on the vacuum with the primary fields at zero and infinity respectively
 \be
 |\alpha\rangle =V_\alpha (0) \, |0\rangle     \qquad ~~~~~~~~~~~~~~~~~
 \langle \alpha | = \lim_{z\rightarrow\infty}z^{2\Delta(\alpha)}\langle 0 | \,V_\alpha (z)
 \ee
Any correlator of the composite fields $V_{\alpha}$ factorizes into the product of a Heisenberg and a Virasoro part. The Heisenberg part, which are just free bosons, is easy to
compute and it reads
\be
 \langle   {\cal V}^{\rm heis}_{\alpha_1} (z_1)  \,  {\cal V}^{\rm heis}_{\alpha_2} (z_2)  \,  {\cal V}^{\rm heis}_{\alpha_3} (z_3)  \,
   {\cal V}^{\rm heis}_{\alpha_4} (z_4)  \,  \rangle=\left(1-\frac{z_3}{z_2} \right)^{2\alpha_2(Q-\alpha_3)}
   \label{4heis}
 \ee
 Consider the remaining Virasoro part of the four-point correlator
 \be
 {\cal G}_{\rm vir}\left(\alpha_i,\alpha | z_i \right)=z_1^{{2\Delta(\alpha_1)}}\langle\langle   {\cal V}^{\rm vir}_{\alpha_1} (z_1)  \,  {\cal V}^{\rm vir}_{\alpha_2} (z_2)  \,  {\cal V}^{\rm vir}_{\alpha_3} (z_3)  \,
   {\cal V}^{\rm vir}_{\alpha_4} (z_4)  \, \rangle\rangle_{\alpha}
 \ee
where by $\langle\langle ~\rangle\rangle_{\alpha}$ I denote the four-point conformal block involving the exchange of a state of conformal dimension
   \be
   \Delta=\Delta(\alpha)
   \ee
and the factor $z_1^{{2\Delta(\alpha_1)}}$ is included to guarantee a finite limit at $z_1 \rightarrow \infty $.
The fact that the 4-point correlator  depends non-trivially only on the cross ratio follows from conformal invariance, so without loss of generality  three points can be fixed: $z_1=\infty$, $z_2=1$ and
$z_4=0$ The resulting function of a single variable $z\equiv z_3$ is denoted
as $ {\cal G}_{vir}(\alpha_i,\alpha|z)$ or simply as $ {\cal G}_{vir}$
if it is clear from the context, what are the argument and the parameters.
Derivatives $ \partial_{z_i} {\cal G}_{\rm vir}$ of the correlator can be also written in terms of derivative with respect to $z$. For the choice above one finds
\cite{Fucito:2013fba}
  \bea
 \partial_{z_1}{\cal G}_{\rm vir}&=& 0 \qquad~~~\qquad   \partial_{z_2} {\cal G}_{\rm vir}=(-z \partial_z
+2 \Delta_1-\delta)\,{\cal G}_{vir}    \nn\\
  \partial_{z_3} {\cal G}_{\rm vir} &=& \partial_z \,{\cal G}_{vir}  \quad \qquad
  \partial_{z_4}{\cal G}_{\rm vir}=((z-1) \partial_z+\delta-2 \Delta_1)\,
{\cal G}_{vir}
 \eea
  with $\delta=\sum_{i=1}^4 \Delta_i$ and $\Delta_i=\Delta(\alpha_i)$.
 Including also the contribution of the Heisenberg sector  one finds the conformal block
 \bea
 {\cal G}(\alpha_i,\alpha|z)&\equiv &\langle\langle \alpha_1 | V_{\alpha_2}(1) \, V_{\alpha_3}(z) \, |\alpha_4 \rangle\rangle_{\alpha}
 = (1-z)^{2\alpha_2(Q-\alpha_3)}{\cal G}_{vir}(\alpha_i,\alpha|z)
\label{confblock}
 \eea
    The ``physical"  correlator can be written
   as the integral of the modulus square of the conformal block (\ref{confblock})
   \be
   G(\alpha_i|z)=\int {d\alpha \over 2\pi} \, C_{\alpha_1 \alpha_2 \alpha}\, C_{\alpha\alpha_3 \alpha_4}\,
   |{\cal G}(\alpha_i,\alpha|z) |^2
\label{fullcorr}
   \ee
   where $C_{\alpha_1\alpha_2\alpha}$ are the Liouville structure
   constants \cite{Zamolodchikov:1995aa,Dorn:1994xn}

\subsubsection{Integrals of motion  }
\label{IM}

 Being an integrable system, the Liouville  theory admits the existence of an infinite set of mutually commuting operators or integrals of motion. Explicitly
the first three such integrals are  \cite{Alba:2010qc}
\bea
 && I_2 = L_0-\ft{c}{24}+2 \sum_{k=1}^\infty a_{-k}\, a_k\nn\\
  && I_3 =   \sum_{k=-\infty,k\ne 0}^{\infty} a_{-k}\, L_k+ 2\, \ii\, Q \sum_{k=1}^\infty k\, a_{-k}\, a_k+
  \ft13 \sum_{i+j+k=0} a_{i}\,a_j\, a_{k}  \\
  && I_4 = 2  \sum_{k=1 }^\infty L_{-k}\, L_k+L_0^2-\ft{c+2}{12}L_0+6 \sum_{k=-\infty,k\ne 0}^{\infty}\sum_{i+j=k} L_{-k}\, a_i\, a_j
+12(L_0-\ft{c}{24})\sum_{k=1}^\infty  a_{-k}\, a_k\nn\\
  &&+6 \ii Q \sum_{k=-\infty,k\ne 0}^{\infty}|k|  a_{-k}\, L_k +2(1-5\, Q^2)\sum_{k=1}^\infty k^2\, a_{-k}\, a_k+
  6 \ii Q\sum_{i+j+k=0} |k| a_{i}\,a_j\, a_{k}\nn\\
&& +\sum_{i+j+k+l=0} :\, a_{i}\,a_j\, a_k \,a_{l} \,: \nn
 \label{intmotion}
\eea
These operators can be inserted inside the four-point correlators and the corresponding conformal blocks.  I define
 \be
 {\cal G}_n(\alpha_i,\alpha|z)=\langle\langle \alpha_1 | V_{\alpha_2}(1) \,I_n\,  V_{\alpha_3}(z) \, |\alpha_4 \rangle\rangle_{\alpha}
 \ee
 To compute  ${\cal G}_n$, I can use the commutation relations (\ref{commrel}) to bring creator and annihilator operators to the left and right sides of the correlation respectively. For instance \footnote{One
should be careful to take into account that the commutators $[L_n,V_{\alpha}]$ produce derivatives only of the Virasoro part of the
 composite field $V_{\alpha}$.
}
  {\small
 \bea
  {\cal G}_2 &=& \langle\langle \alpha_1 | V_{\alpha_2}(1) \,  [L_0,V_{\alpha_3}(z)] \, |\alpha_4 \rangle\rangle_{\alpha} +2 \sum_{k=1}^\infty
  \langle\langle \alpha_1 | [V_{\alpha_2}(1),a_{-k}] \,  [a_k,V_{\alpha_3}(z)] \, |\alpha_4 \rangle\rangle_{\alpha} +(\Delta_4-\ft{c}{24} ){\cal G}\nn\\
  &=& \left( z\partial_z+ {2\alpha_2(Q-\alpha_3) z\over 1-z}+\Delta_3+\Delta_4-\ft{c}{24}- {2\alpha_2(Q-\alpha_3) z\over 1-z} \right) {\cal G}\nn\\
&=&\left( z\partial_z+\Delta_3+\Delta_4-\ft{c}{24}\right)
{\cal G}
 \eea
 }
 Similarly for ${\cal G}_3$ one finds
 {\small
 \bea
 {\cal G}_3
  &=&   \ii \sum_{k=1}^\infty     z^k  \left[ z (Q+\alpha_2-\alpha_3)  \left(  \partial_{z} +{2 \alpha_2 (Q-\alpha_3)\over 1-z }\right)
  +(k+1) \alpha_2 \Delta_3 +(k-1)(Q-\alpha_3)   \Delta_2 \right] {\cal G} \nn\\
  && \ii \left[ -2Q \alpha_2(Q-\alpha_3)  \sum_{k=1}^\infty     z^k
   +\alpha_2 (Q-\alpha_3) (Q+\alpha_2-\alpha_3)  \sum_{i,j=1}^\infty z^{i+j}     \right] {\cal G} \nn\\
   &=& {\ii \, z \over 1-z} \left[  (Q+\alpha_2-\alpha_3)  \, z \partial_z+(Q-\alpha_3)(\Delta_2+\Delta_3+\Delta_4-\Delta_1)-2\alpha_2(Q-\alpha_3)^2) \right]
{\cal G}
 \eea
 }
   Proceeding in this way one can write ${\cal G}_n$ in terms of
 ${\cal G}$ and their $z$-derivatives, or as differential operators acting on ${\cal G}$. Moreover
\be
{\cal G}_n(\alpha_i,\alpha|z)=  {\cal L}_n \, {\cal G}(\alpha_i,\alpha|z)
\label{gndef}\ee
The explicit form of the differential operators on the variable $z$, the ${\cal L}_n$'s, up to $n=4$, can be found in \cite{Fucito:2015ofa}.

\subsection{The gauge/CFT dictionary}

The AGT correspondence relates the four-point conformal block of the Liouville theory to the partition function of  the ${\cal N}=2$ supersymmetric $SU(2)$ gauge theory with four fundamentals.  The gauge coupling parameter $q=e^{2\pi {\rm i} \tau}$  is identified with the harmonic ratio $z$ parametrizing the positions of vertex insertions. The gauge theory masses $m_u, \bar m_{2+u}$  are related to the conformal dimensions of the vertex insertions in the CFT.
To achieve a full symmetry with respect to the exchange of the four masses we make
the replacements  $m_3\to \bar{m}_3+\epsilon $, $m_4\to \bar{m}_4+\epsilon $.
The vacuum expectation value $a$ for the scalar field at infinity  parametrizes the dimension of the exchanged state.
The squeezing parameter $\epsilon_1/\epsilon_2$ characterizing the $\Omega$ gravitational background parametrizes the central charge of the CFT. The full dictionary
is given by \cite{Alday:2009aq}
\bea
\alpha_1 &=&  \ft{\epsilon}{2}+\ft12(\bar{m}_1-\bar{m}_2)
 \qquad   \alpha_2 =  -\ft12 (\bar{m}_1+\bar{m}_2 ) \nn\\
  \alpha_3 &=& \epsilon+\ft12 ( \bar m_3+\bar m_4)  \qquad \alpha_4= \ft{\epsilon}{2}+  \ft12( \bar m_3- \bar m_4)
 \nn\\
  \alpha &=&   \ft{\epsilon}{2}+  a  \qquad \epsilon=\epsilon_1+\epsilon_2=Q\qquad
  \epsilon_1= b \qquad  \epsilon_2=b^{-1}  \qquad z=q     \label{agt}
   \eea
  The instanton partition function of the gauge theory on $
\R^4$ is  related to the conformal block ${\cal G}(\alpha_i,\alpha|q)$ via
 \be
 Z^{U(2)}_{\rm tree+inst}  (a,m_i,q) = q^{-a^2  }  \, Z^{U(2)}_{\rm inst} (a,m_i,q) =q^{ -{Q^2 \over 4} +\Delta_3 +\Delta_4} \,
 {\cal G}(\alpha,\alpha_i,q)
\label{ZU2}
 \ee
  with $Z_{\rm inst}\sim 1$ and ${\cal G}\sim
   q^{\Delta-\Delta_3 -\Delta_4} $ for small $q$.
 On the other hand,  the Virasoro conformal block is related to the SU(2) partition function  via
  \be
{\cal G}_{\rm vir}(\alpha,\alpha_i, q)   = q^{ \Delta-\Delta_3 -\Delta_4}  \,  Z^{SU(2)}_{\rm inst} (a,m_i q) = q^{ \Delta-\Delta_3 -\Delta_4}  \,  (1-q)^{-2\alpha_2(Q-\alpha_3)} \,
Z_{\rm inst}^{U(2)} (a,m_i q)
 \label{ZSU2} \ee
  with  the extra factor canceling the $U(1)$  contribution (\ref{4heis})  arising from the Heisenberg CFT field.
The full four-point correlator (\ref{fullcorr})
 is then identified with the gauge partition function on the sphere via
\be
G(\alpha_i,q) =  Z^{U(2)}_{S^4} (m_i, q)
\ee

 \subsection{  The gauge theory side }

 It is known \cite{Nekrasov:2009rc,Bonelli:2009zp,Bonelli:2014iza} that the integrals of motion (\ref{intmotion}) can be put in relation to the chiral correlators $\left\langle {\rm tr}\, \tilde{\varphi}^J \right\rangle$.
In this section I translate the formulae in the previous section in terms of the gauge theory variables to find that the chiral correlators in the undeformed $U(2)$ gauge theory
can be expressed  in terms of q-derivatives of the partition function $Z$. This leads to chiral ring type relations valid at all-instanton orders for a finite $\Omega$-background.

\subsubsection{Chiral relations: $\epsilon_1,\epsilon_2$ finite}

 The result (\ref{gndef}) can be translated into chiral correlators using the identification
  \bea
 \langle  {\rm tr} \tilde{\varphi}^2 \rangle &=&  -2\,  \frac{{\cal G}_2(q)}{{\cal G}}- \ft{1}{12}          \nn\\
    \langle  {\rm tr} \tilde{\varphi}^3 \rangle &=& 6 \,\ii \,  \frac{{\cal G}_3(q)}{{\cal G}}\\
\langle  {\rm tr} \tilde{\varphi}^4 \rangle &=&2 h^4\frac{{\cal G}_4(q)}{{\cal G}}-\frac{h^2}{4}\langle  {\rm tr} \tilde{\varphi}^2 \rangle+\frac{\epsilon^2(h^2+\epsilon^2)}{8}\nn
 \eea
 where
 \be
   \epsilon= \epsilon_1+\epsilon_2 \qquad ~~~~~~~~    h^2= \epsilon_1 \epsilon_2
   \ee
Using the AGT dictionary ( \ref{agt}), (\ref{ZU2}) leads to
   {\small
    \bea
     \langle  {\rm tr} \tilde{\varphi}^2 \rangle &=&   - 2 \, h^2 \,   {q\,\partial_q Z  \over   Z } \nn\\
    \langle  {\rm tr} \tilde{\varphi}^3 \rangle &=& \frac{ 3 \, q }{1-q}
      \, \left ( - h^2 \,  {M}_{1}   {q\,\partial_q Z \over   Z  }  +{M}_{3}  \right )\nn\\
       \langle  {\rm tr} \tilde{\varphi}^4 \rangle &=&
       \frac{  2 \, q  }{{\left ( 1-q\right ) }^{2}}    \, \left ( 2  \, \epsilon\, {M}_{3} +2 M_4  +2 \, q \, \left ( {M}_{1} \, {M}_{3} -M_4\right )   +
     h^4  \,q(1-q^2)  { \partial^2_q Z  \over   Z  }   \right. \nn\\
     && \left.   +h^2 \,\left[ h^2-2 q (\epsilon M_1+M_2)+q^2(-h^2+2 M_2-2 M_1^2)  \right]{ \partial_q Z  \over   Z  } \right) \label{recurr}
           \eea
   }
with $Z=Z^{U(2)}_{\rm one-loop} Z^{U(2)}_{\rm inst+tree}$ and
   \be
   M_1=-\sum_{i=1}^4 \bar{m}_i  \quad   M_2=\sum_{i<j}^4 \bar{m}_i \bar{m}_j \quad
    M_3=-\sum_{i < j< k}^4 \bar{m}_i \bar{m}_j \bar{m}_k  \quad M_4=\,
\bar{m}_1 \bar{m}_2 \bar{m}_3 \bar{m}_4
   \ee
   Notice that the last two equations of (\ref{recurr}) can be rewritten in the form
 \begin{eqnarray}
  \langle  {\rm tr} \tilde{\varphi}^3 \rangle &=& {3\,q\over 1-q} \left( \ft12 \,   \langle  {\rm tr} \tilde{\varphi}^2 \rangle\, M_1+  M_3\right) \label{recurr22} \\
  \langle  {\rm tr} \tilde{\varphi}^4 \rangle &=& {(1+q)\,   \over 2(1-q)}  \langle  {\rm tr} \tilde{\varphi}^2 \rangle^2
  + {2q\over (1-q)^2}
  \left(   M_2(1-q) +q M_1^2 +\epsilon M_1 \right)  \langle  {\rm tr} \tilde{\varphi}^2 \rangle
  \nn\\
 &&
 -h^2 {1+q\over 1-q}\, q\partial_q \langle  {\rm tr} \tilde{\varphi}^2 \rangle
   + {4q\over (1-q)^2}
  \left(   M_4(1-q) +q M_1 M_3 +\epsilon M_3 \right) \nn
\end{eqnarray}
which shows that in a finite $\Omega$-background, chiral correlators can be written in terms of  both $\langle  {\rm tr} \tilde{\varphi}^2 \rangle$ and its derivatives.
The chiral ring equations (\ref{recurr22}) generalize those found in \cite{Cachazo:2002ry} to the case of finite $\epsilon_1,\epsilon_2$.
The relations (\ref{recurr}) can be checked against a microscopic instanton computation showing agreement \cite{Fucito:2015ofa}.

\vskip 1.5cm
\noindent {\large {\bf Acknowledgments}}
\vskip 0.2cm
Francesco Fucito wants to thanks the organizers of the conference for the invitation which gave him the opportunity
to spend some time in company of some old friends, collaborators, ex-students discussing physics, mathematics and some less academic issues.

\providecommand{\href}[2]{#2}\begingroup\raggedright\endgroup

\end{document}